\documentclass[conference, letterpaper]{IEEEtran}
\pdfoutput=1

\newcommand{\isaccepted}{}
\newcommand{\isarxiv}{}

\usepackage[utf8]{inputenc}

\ifCLASSOPTIONcompsoc
\usepackage[nocompress]{cite}
\else
\usepackage{cite}
\fi

\usepackage{lipsum}
\usepackage{graphicx}

\usepackage{url}

\usepackage{amsmath}
\usepackage{amssymb}
\usepackage{bm}
\usepackage{mathtools}
\usepackage{tabularx}
\newcolumntype{T}{>{\ttfamily}X}

\usepackage[operators]{cryptocode}
\usepackage[misc,geometry]{ifsym}

\ifdefined\isarxiv%
\usepackage{hyperref}
\hypersetup{
    final,
    colorlinks=true, breaklinks=true, bookmarks=false,bookmarksnumbered,
    urlcolor=EricssonBlue, linkcolor=EricssonBlack, citecolor=EricssonGreen, % Link colors
    pdftitle={mmWave Beam Selection in Analog Beamforming using Personalized Federated Learning}, % PDF title
    pdfauthor={Martin Isaksson, Filippo Vannella, David Sandberg, Rickard Cöster}, % PDF Author
    pdfkeywords={}, % PDF Keywords
    pdfcreator={PdfLaTeX}, % PDF Creator
}
\fi
\makeatletter
\g@addto@macro\UrlBreaks{\do\-\do\/}
\makeatother

\usepackage{tikz}
\usepackage[export]{adjustbox}
\usepackage{balance}
\usepackage{algorithm}

\usepackage{pgfplots}

\usepackage[noend]{algpseudocode}         % https://en.wikibooks.org/wiki/LaTeX/Algorithms#Typesetting_using_the_algorithmic_package
\usepackage{microtype}

\usepackage[capitalise]{cleveref}  % Clever references to stuff.

\usepackage{booktabs}              % Nicer tables
\usepackage[round-mode=places,detect-weight=true,round-pad = false]{siunitx}  % 
\DeclareSIUnit{\belmilliwatt}{Bm}
\DeclareSIUnit{\dBm}{\deci\belmilliwatt}
\DeclareSIUnit\px{px}

\usepackage{multirow}
\usepackage{array}
\usepackage{arydshln}

\microtypecontext{spacing=nonfrench}
\ifdefined\isarxiv%
\microtypesetup{
      activate={true, nocompatibility},
    protrusion=true,
     final,
     tracking=false,
     kerning=true,
     spacing=true,
     factor=1100,
     shrink=30
}
\SetTracking{encoding={*}, shape=sc}{40}
\else
\microtypesetup{
      activate={true, nocompatibility},
      protrusion=false,
}
\fi
\usepackage[graph, grays]{ericolors}
\usepackage{xspace}
\newcommand\hl[1]{%
    \tikz[baseline,%
      outer sep=-15pt, inner sep = 2pt%
    ]%
    \node[rounded corners,rectangle,fill=EricssonBlue4,anchor=text]{#1\xspace};%
}%
\newcommand\hlf[1]{%
    \tikz[baseline,%
      outer sep=-15pt, inner sep = 2pt%
    ]%
   \node[rounded corners,rectangle,fill=EricssonGreen4,anchor=text]{#1\xspace};%
}%

\usetikzlibrary{quotes}
\usetikzlibrary{patterns,shapes.arrows, arrows.meta, fit, calc, intersections}
\usetikzlibrary{shapes, decorations, arrows, angles}
\usetikzlibrary{3d, fit, backgrounds, decorations.text}
\usetikzlibrary{positioning, shapes.symbols, tikzmark, decorations.markings}
\usetikzlibrary{decorations.pathreplacing, calligraphy, 3d, perspective}

\tikzset{>=latex}
\tikzset{mydeco/.style={pic actions/.append code=\tikzset{postaction=decorate}}}

\pgfplotsset{compat=newest,
    width=6cm,
    height=3.5cm,
    scale only axis=true,
    max space between ticks=25pt,
    try min ticks=5,
    every axis/.style={
        axis y line=left,
        axis x line=bottom,
        axis line style={thick,->,>=latex, shorten >=-.4cm},
        mark size=2pt,
        label style={font=\large},
        tick label style={font=\large}
    },
    every axis plot/.append style={thick},
    tick style={EricssonBlack, thick},
    colorbar style={at={(1.25, 1)}}
}

\usepgfplotslibrary{groupplots}

\tikzset{
    semithick/.style={line width=0.8pt},
}

\newcommand{\R}{\mathbb{R}}
\newcommand{\N}{\mathbb{N}}

\usepackage[nomain,acronym,section=subsection,symbols, numberedsection]{glossaries-extra}
\setabbreviationstyle[acronym]{long-short}

\ifdefined\isarxiv%

    \makeglossaries
\fi%

\DeclareMathOperator{\Prob}{\mathcal{P}}
\DeclareMathOperator{\sgn}{sgn}

\renewcommand{\epsilon}{\varepsilon}

\newcommand{\flconstfnt}[1]{\ensuremath{\textsc{#1}}}

\def\etal.{et\penalty50\ al.}
\newcommand{\di}{{\mathrm{i}\mkern1mu}}

\providecommand{\tightlist}{%
\setlength{\itemsep}{0pt}\setlength{\parskip}{0pt}\setlength{\parindent}{0pt}}

\let\tightlist\relax
\def\tightlist{}

\renewcommand\fbox{\fcolorbox{EricssonGreen}{white}}

\newcommand\copyrighttext{%
  \footnotesize \textcopyright \the\year{} IEEE. Personal use of this material is permitted. Permission from IEEE must be obtained for all other uses, including reprinting/republishing this material for advertising or promotional purposes, collecting new collected works for resale or redistribution to servers or lists, or reuse of any copyrighted component of this work in other works.}

\newcommand\copyrightnotice{%
\begin{tikzpicture}[remember picture,overlay]
\node[anchor=south,yshift=10pt] at (current page.south) {\fbox{\parbox{\dimexpr\textwidth-\fboxsep-\fboxrule\relax}{\copyrighttext}}};
\end{tikzpicture}%
}

\newacronym{3GPP}{3GPP}{Third Generation Partnership Project}
\newacronym{5G}{5G}{Fifth-generation technology standard}
\newacronym{AWGN}{AWGN}{Additive White Gaussian Noise}
\newacronym{BS}{BS}{Base Station}
\newacronym{ANN}{ANN}{Artificial Neural Network}
\newacronym{API}{API}{Application Programming Interface}
\newacronym{CA}{CA}{Certificate Authority}
\newacronym{CDF}{CDF}{Cumulative Distribution Function}
\newacronym{CIFAR}{CIFAR}{Canadian Institute For Advanced Research}
\newacronym{CNN}{CNN}{Convolutional Neural Network}
\newacronym{DEF}{DEF}{Data Expansion Factor}
\newacronym{DFT}{DFT}{Discrete Fourier Transform}
\newglossaryentry{DL}{
    name={DL},
    description={downlink, signal coming from a cell tower to your mobile device},
    first={downlink (DL)},
    long={downlink}
}

\newacronym{DoS}{DoS}{Denial-of-service}
\newacronym{ECDF}{ECDF}{Empirical Cumulative Density Function}
\newacronym{ECIES}{ECIES}{Elliptic Curve Integrated Encryption Scheme}
\newacronym{FedAdam}{\textsc{FedAdam}}{Federated Adam}
\newacronym{FedAdagrad}{\textsc{FedAdagrad}}{Federated Adagrad}
\newacronym{FedAvg}{\textsc{FedAvg}}{Federated Averaging}
\newacronym{FedLion}{\textsc{FedLion}}{Federated Evo\textbf{L}ved S\textbf{i}gn M\textbf{o}me\textbf{n}tum}
\newacronym{FedYogi}{\textsc{FedYogi}}{Federated Yogi}
\newacronym{FEMNIST}{FEMNIST}{Federated Extended MNIST}
\newacronym{FL}{FL}{Federated Learning}
\newacronym{FQDN}{FQDN}{Fully Qualified Domain Name}
\newacronym{GPU}{GPU}{Graphics Processing Unit}
\newacronym{GoB}{GoB}{Grid-of-beams}
\newacronym{IID}{IID}{Independent and Identically Distributed}
\newacronym{IFCA}{IFCA}{Iterative Federated Clustering Algorithm}
\newacronym{IoT}{IoT}{Internet of Things}
\newacronym{IP}{IP}{Internet Protocol}
\newacronym{KDF}{KDF}{Key Derivation Function}
\newacronym{KPC}{KPC}{Kronecker-product based codebook}
\newacronym{KPI}{KPI}{Key Performance Indicator}
\newacronym{KTH}{KTH}{KTH Royal Institute of Technology}
\newacronym{Lion}{Lion}{Evo\textbf{L}ved S\textbf{i}gn M\textbf{o}me\textbf{n}tum}
\newacronym{MAB}{MAB}{Multi-Armed Bandit}
\newacronym{MAC}{MAC}{Message Authentication Code}
\newacronym{ML}{ML}{Machine Learning}
\newglossaryentry{mmWave}{
    name={mmWave},
    description={the band of spectrum with wavelengths between 10 millimeters (30 GHz) and 1 millimeter (300 GHz)},
    first={Millimeter Wave (mmWave)},
    long={mmWave},
    short={mmWave}
}

\newacronym{MoE}{MoE}{Mixture of Experts}
\newacronym{MPC}{MPC}{Multi-Party Computation}
\newacronym{NR}{NR}{New Radio}
\newacronym{PKI}{PKI}{Public Key Infrastructure}
\newacronym{PLMN}{PLMN}{Public Land Mobile Network}
\newacronym{PRF}{PRF}{Pseudo Random Function}
\newacronym{PRG}{PRG}{Pseudo Random Generator}
\newacronym{PS}{PS}{Public Safety}
\newacronym{PRNG}{PRNG}{Pseudo Random Number Generator}
\newacronym{QoE}{QoE}{Quality of Experience}
\newacronym{RAN}{RAN}{Radio Access Network}
\newacronym{RF}{RF}{Radio Frequency}
\newacronym{RSA}{RSA}{Rivest-Shamir-Adleman}
\newacronym{RSRP}{RSRP}{Reference Signal Received Power}
\newacronym{SNIC}{SNIC}{Swedish National Infrastructure for Computing}
\newacronym{SAR}{SAR}{Search and Rescue}
\newacronym{SGD}{SGD}{Stochastic Gradient Descent}
\newacronym{SNR}{SNR}{Signal-to-noise ratio}
\newacronym{TCP}{TCP}{Transmission Control Protocol}
\newacronym{TLS}{TLS}{Transport Layer Security}
\newacronym{UDP}{UDP}{User Datagram Protocol}
\newacronym{UE}{UE}{User Equipment}

\newglossaryentry{UL}{
    name={UL},
    description={uplink, signal leaving your mobile device and going back to a 
    cell tower},
    first={uplink (UL)},
    long={uplink}
}

\newacronym{UPF}{UPF}{User Plane Function}
\newacronym{UAV}{UAV}{Unmanned Aerial Vehicle}
\newacronym{USV}{USV}{Unmanned Surface Vehicle}
\newacronym{UCB}{UCB}{Upper Confidence Bound}
\newacronym{ULA}{ULA}{Uniform Linear Array}
\newacronym{UPA}{UPA}{Uniform Planar Array}
\newacronym{UUID}{UUID}{Universally Unique Identifier}
\newacronym{WARA-PS}{WARA-PS}{WASP Autonomous Research Arenas --- Public Safety}
\newacronym{WARA}{WARA}{WASP Autonomous Research Arenas}
\newacronym{WASP}{WASP}{Wallenberg AI, Autonomous Systems and Software Program}

\newacronym{AF}{AF}{Application Function}
\newacronym{NF}{\text{NF}}{Network Function}
\newacronym{NRF}{NRF}{NF Repository Function}
\newacronym{NWDA}{NWDA}{Network Data Analytics}
\newacronym{NWDAF}{NWDAF}{Network Data Analytics Function}
\newacronym{AMF}{AMF}{Access and Mobility Management Function}
\newacronym{AUSF}{AUSF}{AUthentication Server Function}
\newacronym{OAM}{OAM}{Operation, Administration, and Maintenance}
\newacronym{SBA}{SBA}{Service Based Architecture}

\newglossaryentry{sub6}{
    name={Sub-6GHz},
    sort={sub6ghz},
    description={frequency bands under 6 GHz},
    first={Sub-6GHz},
    long={Sub-6GHz}
}

\newglossaryentry{Htrk}{name={\ensuremath{\bm{H}_k^{t,r}}}, sort={Htrk}, description={frequency-domain channel matrix for subcarrier \(k\)}, type={symbols}}

\newglossaryentry{Htr}{name={\ensuremath{\bm{H}^{t,r}}}, sort={Htr}, description={frequency-domain channel matrix}, type={symbols}}

\newglossaryentry{eta}{name={\ensuremath{\eta}}, sort={eta}, description={Learning rate}, type={symbols}}

\newglossaryentry{beta1}{name={\ensuremath{\beta_1}}, sort={beta}, description={Lion momentum hyper-parameter}, type={symbols}}
\newglossaryentry{beta2}{name={\ensuremath{\beta_2}}, sort={beta}, description={Lion momentum hyper-parameter}, type={symbols}}

\newglossaryentry{lambda}{name={\ensuremath{\lambda}}, sort={lambda}, description={weight decay}, type={symbols}}
\newglossaryentry{subcarrierbandwidth}{name={\ensuremath{B}}, sort={B}, description={Subcarrier bandwidth}, type={symbols}}

\newglossaryentry{C}{name={\ensuremath{C}}, sort={C}, description={Fraction of clients selected}, type={symbols}}
\newglossaryentry{E}{name={\ensuremath{E}}, sort={E}, description={local epochs}, type={symbols}}

\newglossaryentry{eps}{name={\ensuremath{\varepsilon}}, sort={eps}, description={\(\varepsilon\)-greedy parameter}, type={symbols}}

\newglossaryentry{t}{name={\ensuremath{t}},sort={t},description={time in communication rounds},type={symbols}}
\newglossaryentry{flf}{name={\ensuremath{\mathcal{L}}},sort={L},description={Estimated total loss},type={symbols}}
\newglossaryentry{fl}{name={\ensuremath{f_l}},sort={fl},description={Local model},type={symbols}}
\newglossaryentry{flk}{name={\ensuremath{f^k_l}},sort={flk},description={Local model for client $k$},type={symbols}}
\newglossaryentry{flkp}{name={\ensuremath{f^{k'}_l}},sort={flkp},description={Local model for client $k'$},type={symbols}}
\newglossaryentry{fgj}{name={\ensuremath{f^j_g}},sort={fgj},description={Cluster model with index $j$},type={symbols}}
\newglossaryentry{fg}{name={\ensuremath{f_g}},sort={fgj},description={Global model},type={symbols}}

\newglossaryentry{F}{name={\ensuremath{\mathcal{F}}},sort={F},description={Code book for downlink beamforming},type={symbols}}

\newglossaryentry{fi}{name={\ensuremath{\bm{\mathrm{f}}_i}},sort={fi},description={Code book index $i$, or downlink beamforming vector},type={symbols}}
\newglossaryentry{f}{name={\ensuremath{\bm{\mathrm{f}}}},sort={f},description={Generic code book, or downlink beamforming vector},type={symbols}}
\newglossaryentry{fml}{name={\ensuremath{\mathcal{F}_{m,l}}},sort={f},description={Generic code book, or downlink beamforming vector},type={symbols}}

\newglossaryentry{nobeams}{name={\ensuremath{N}},sort={N},description={Number of beams},type={symbols}}

\newglossaryentry{Dn}{name={\ensuremath{D_n}},sort={dn},description={Dataset of length $n$},type={symbols}}
\newglossaryentry{p}{name={\ensuremath{p}}, sort={p}, description={majority class fraction}, type={symbols}}

\newglossaryentry{P}{name={\ensuremath{\Prob}}, sort={P}, description={probability}, type={symbols}}
\newglossaryentry{dltxpow}{name={\ensuremath{P}}, sort={P}, description={Downlink transmit power in Watt}, type={symbols}}

\newglossaryentry{noisepower}{name={\ensuremath{\sigma^2}}, sort={s}, description={Noise power per subcarrier}, type={symbols}}

\newglossaryentry{w}{name={\ensuremath{\bm{w}}}, sort={w}, description={model parameters}, type={symbols}}
\newglossaryentry{wk}{name={\ensuremath{\bm{w}_k}}, sort={wk}, description={Local model parameters}, type={symbols}}
\newglossaryentry{whk}{name={\ensuremath{\bm{w}^k_h}}, sort={whk}, description={Gating model parameters for client $k$}, type={symbols}}
\newglossaryentry{wlk}{name={\ensuremath{\bm{w}^k_l}}, sort={wlk}, description={Local model parameters for client $k$}, type={symbols}}
\newglossaryentry{wlt}{name={\ensuremath{\bm{w}_l(t)}}, sort={wlt}, description={Local model parameters at time $t$}, type={symbols}}
\newglossaryentry{wkt}{name={\ensuremath{\bm{w}^k(t)}}, sort={wkt}, description={Global model parameters for global model at time $t$}, type={symbols}}

\newglossaryentry{wkt1}{name={\ensuremath{\bm{w}^k(t+1)}}, sort={wlt}, description={Global model para meters for global model at time $t+1$}, type={symbols}}
\newglossaryentry{wg}{name={\ensuremath{\bm{w}_{g}}}, sort={wg}, description={Global model parameters}, type={symbols}}
\newglossaryentry{wgt}{name={\ensuremath{\bm{w}_{g}(t)}}, sort={wgt}, description={Global model parameters at time $t$}, type={symbols}}
\newglossaryentry{wg0}{name={\ensuremath{\bm{w}_{g}(0)}}, sort={wgt}, description={Global model parameters at time 0}, type={symbols}}
\newglossaryentry{wgt1}{name={\ensuremath{\bm{w}_{g}(t+1)}}, sort={wgt}, description={Global model parameters at time $t+1$}, type={symbols}}

\newglossaryentry{wgj}{name={\ensuremath{\bm{w}^j_g}}, sort={wgj}, description={Cluster model $j$ parameters}, type={symbols}}
\newglossaryentry{wgjt}{name={\ensuremath{\bm{w}^j_g(t)}}, sort={wgjt}, description={Cluster model $j$ parameters at time $t$}, type={symbols}}
\newglossaryentry{wgjht}{name={\ensuremath{\bm{w}^{\hat{j}}_g(t)}}, sort={wgjt}, description={Cluster model $\hat{j}$ parameters at time $t$}, type={symbols}}

\newglossaryentry{wgt0}{name={\ensuremath{\bm{w}_g(0)}}, sort={wgt0}, description={Global model parameters at time $0$}, type={symbols}}

\newglossaryentry{wgjt0}{name={\ensuremath{\bm{w}^j_g(0)}}, sort={wgjt0}, description={Cluster model $j$ parameters at time $0$}, type={symbols}}
\newglossaryentry{wgjt1}{name={\ensuremath{\bm{w}^j_g(t+1)}}, sort={wgjt1}, description={Cluster model $j$ parameters at time $t+1$}, type={symbols}}

\newglossaryentry{gl}{name={\ensuremath{g_l}},sort={gl},description={Gate model weight for local model},type={symbols}}
\newglossaryentry{glk}{name={\ensuremath{g_l^k}},sort={glk},description={Gate model weight for local model for client $k$},type={symbols}}
\newglossaryentry{g}{name={\ensuremath{g}},sort={g},description={Gate model weight},type={symbols}}
\newglossaryentry{gjk}{name={\ensuremath{g_j^k}},sort={gjk},description={Gate model weight for cluster model $j$ and client $k$},type={symbols}}

\newglossaryentry{h}{name={\ensuremath{f_h}},sort={h},description={Gate model function},type={symbols}}
\newglossaryentry{hk}{name={\ensuremath{f_h^k}},sort={hk},description={Gate model for client $k$},type={symbols}}
\newglossaryentry{hkp}{name={\ensuremath{f_h^{k'}}},sort={hk},description={Gate model for client $k$},type={symbols}}

\newglossaryentry{hul}{name={\ensuremath{\bm{\mathrm{h}}^{\text{UL}}}},sort={hul},description={Uplink channel estimate},type={symbols}}
\newglossaryentry{huli}{name={\ensuremath{\bm{\mathrm{h}}_i^{\text{UL}}}},sort={huli},description={Uplink channel estimate sample $i$},type={symbols}}

\newglossaryentry{hdl}{name={\ensuremath{\bm{\mathrm{h}}_i^{\text{DL}}}},sort={hdl},description={Downlink channel estimate},type={symbols}}
\newglossaryentry{hdli}{name={\ensuremath{\bm{\mathrm{h}}_i^{\text{DL}}}},sort={hdli},description={Downlink channel estimate sample $i$},type={symbols}}

\newglossaryentry{J}{name={\ensuremath{J}}, sort={J}, description={Number of cluster models}, type={symbols}}

\newglossaryentry{j}{name={\ensuremath{j}}, sort={j}, description={Cluster model index}, type={symbols}}

\newglossaryentry{beamformingindex}{name={\ensuremath{l}}, sort={l}, description={Beamforming vector index}, type={symbols}}

\newglossaryentry{jhat}{name={\ensuremath{\hat{j}}}, sort={jh}, description={Cluster model identity estimate}, type={symbols}}
\newglossaryentry{k}{name={\ensuremath{k}}, sort={k}, description={Client index}, type={symbols}}
\newglossaryentry{K}{name={\ensuremath{K}}, sort={K}, description={Number of clients}, type={symbols}}
\newglossaryentry{kp}{name={\ensuremath{k'}}, sort={kp}, description={Client index, primed}, type={symbols}}

\newglossaryentry{l}{name={\ensuremath{\ell}},sort={l},description={Loss function},type={symbols}}

\newglossaryentry{m}{name={\ensuremath{m}},sort={m},description={Pre-coder weight index},type={symbols}}

\newglossaryentry{M}{name={\ensuremath{M}},sort={M},description={Number of antennas},type={symbols}}
\newglossaryentry{L}{name={\ensuremath{L}},sort={L},description={Number of subcarriers},type={symbols}}

\newglossaryentry{beamangle}{name={\ensuremath{\theta}}, sort={theta}, description={Beam angle}, type={symbols}}

\newglossaryentry{Ks}{name={\ensuremath{K_s}}, sort={ks}, description={Number of selected clients}, type={symbols}}
\newglossaryentry{nk}{name={\ensuremath{n_k}}, sort={nk}, description={Number of data samples for client $k$}, type={symbols}}
\newglossaryentry{nj}{name={\ensuremath{n_j}}, sort={nj}, description={Total number of data samples for cluster $j$ in one iteration.}, type={symbols}}

\newglossaryentry{n}{name={\ensuremath{n}}, sort={n}, description={Total number of data samples}, type={symbols}}

\newglossaryentry{momentum}{name={\ensuremath{\bm{m}}}, sort={m}, description={Momentum}, type={symbols}}

\newglossaryentry{update}{name={\ensuremath{\bm{\Delta}}}, sort={D}, description={Client}, type={symbols}}
\newglossaryentry{clientupdate}{name={\ensuremath{\bm{\Delta}^k}}, sort={D}, description={Client update}, type={symbols}}

\newglossaryentry{x}{name={\ensuremath{\bm{x}}}, sort={x}, description={Features of a data sample}, type={symbols}}
\newglossaryentry{y}{name={\ensuremath{y}}, sort={y}, description={Target of a data sample}, type={symbols}}

\newglossaryentry{i}{name={\ensuremath{i}}, sort={i}, description={Data sample index}, type={symbols}}
\newglossaryentry{xi}{name={\ensuremath{\bm{x}_i}}, sort={xi}, description={Features of data sample $i$}, type={symbols}}
\newglossaryentry{yi}{name={\ensuremath{y_i}}, sort={yi}, description={Ground truth class of data sample $i$}, type={symbols}}
\newglossaryentry{yihat}{name={\ensuremath{\hat{y}_i}}, sort={yihat}, description={Predicted class of data sample $i$}, type={symbols}}
\newglossaryentry{yhatl}{name={\ensuremath{\hat{y}_l}}, sort={yhatl}, description={Estimated target (local)}, type={symbols}}
\newglossaryentry{yhath}{name={\ensuremath{\hat{y}_h}}, sort={yhath}, description={Estimated target (gating)}, type={symbols}}
\newglossaryentry{yhatg}{name={\ensuremath{\hat{y}_g}}, sort={yhatg}, description={Estimated target (gating)}, type={symbols}}
\newglossaryentry{yhatj}{name={\ensuremath{\hat{y}_j}}, sort={yhatj}, description={Estimated target (cluster)}, type={symbols}}

\newglossaryentry{Pk}{name={\ensuremath{P_k}}, sort={Pk}, description={Partition of dataset accessible to client $k$}, type={symbols}}

\newglossaryentry{subcarrierindex}{name={\ensuremath{s}}, sort={s}, description={Subcarrier index}, type={symbols}}

\newglossaryentry{S}{name={\ensuremath{\left\{1,2,\ldots,K\right\}}}, sort={S}, description={Population of clients}, type={symbols}}
\newglossaryentry{St}{name={\ensuremath{S_t}}, sort={St}, description={Selected set of clients at time $t$}, type={symbols}}

\newglossaryentry{Jset}{name={\ensuremath{\left\{1,2,\ldots,J\right\}}}, sort={J}, description={$[J] = \left\{j \in \N^+ \,\middle\vert\, j \leq J\right\}$}, type={symbols}}
\newglossaryentry{Jremset}{name={\ensuremath{\mathcal{J}}}, sort={J}, description={Set of cluster models}, type={symbols}}

\newglossaryentry{Eset}{name={\ensuremath{\left\{1,2,\ldots,E\right\}}}, sort={J}, description={$[E] = \left\{e \in \N^+ \,\middle\vert\, e \leq E\right\}$}, type={symbols}}

\ifCLASSOPTIONcompsoc
\usepackage[caption=false,font=footnotesize,labelfont=sf,textfont=sf]{subfig}
\else
  \usepackage[caption=false,font=footnotesize]{subfig}
\fi

\begin{document}

\title{mmWave Beam Selection in Analog Beamforming\\Using Personalized Federated Learning}

%\ifdefined\isaccepted%
    \author{%
       \IEEEauthorblockN{Martin~Isaksson\IEEEauthorrefmark{1}\IEEEauthorrefmark{2},
       Filippo~Vannella\IEEEauthorrefmark{1}\IEEEauthorrefmark{2},
       David~Sandberg\IEEEauthorrefmark{1} and
       Rickard~Cöster\IEEEauthorrefmark{1}}%
       \IEEEauthorblockA{\IEEEauthorrefmark{1}Ericsson AB, Stockholm, Sweden}
       \IEEEauthorblockA{\IEEEauthorrefmark{2}KTH Royal Institute of Technology, Stockholm, Sweden}%
       \IEEEauthorblockA{Contact: \texttt{martin.isaksson@ericsson.com}}%
    }
%\else%
%    \author{\IEEEauthorblockN{Anonymous Authors}}
%    \usepackage[firstpage]{draftwatermark}
%\fi%

\maketitle
\global\csname @topnum\endcsname 0
\global\csname @botnum\endcsname 0

\begin{abstract}Using analog beamforming in \glsxtrshort{mmWave} frequency bands we can focus
the energy towards a receiver to achieve high throughput. However, this requires
the network to quickly find the best \glsxtrlong{DL} beam configuration in the
face of non-\glsxtrshort{IID} data. We propose a personalized \gls{FL} method to
address this challenge, where we learn a mapping between \glsxtrlong{UL}
\gls{sub6} channel estimates and the best \glsxtrlong{DL} beam in heterogeneous
scenarios with non-\glsxtrshort{IID} characteristics. We also devise
\glsxtrshort{FedLion}, a \gls{FL} implementation of the \glsxtrshort{Lion}
optimization algorithm. Our approach reduces the signaling overhead and
provides superior performance, up to \SI{33.6}{\percent} higher accuracy than a
single \gls{FL} model and \SI{6}{\percent} higher than a local model.\end{abstract}


\glsresetall
\bstctlcite{IEEEexample:BSTcontrol}

\glsresetall
\section{Introduction}
\ifdefined\isarxiv\copyrightnotice\fi
The massive data traffic demands of next generation mobile networks require new
technologies and deployment strategies in high-frequency bands. In particular,
\gls{mmWave} frequency bands are important for \glsxtrshort{5G} mobile
communication due to large available bandwidths which provide low latency, and
high data rates~\cite{ericssonmmwave}. Signals in \gls{mmWave} frequency bands 
are affected by high path loss, and are
sensitive to blockage effects in the environment. Using large antenna arrays,
combined with beamforming techniques, we can focus the energy towards a
receiver. This requires a beam training phase to steer the transmitter beam
towards the receiver, and it is important to do this quickly with low signaling
overhead.

\begin{figure}
    \centering
    \includegraphics[width=0.85\linewidth]{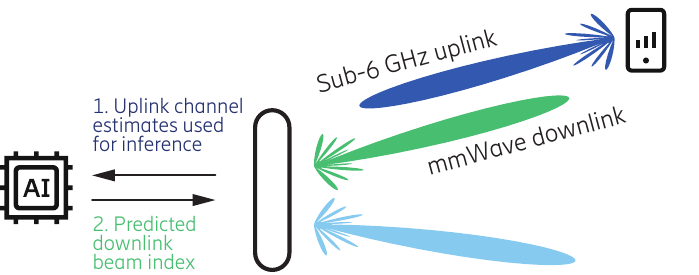}
    \caption{\textbf{Overview of the proposed \gls{mmWave} beam prediction system.} \gls{UL} \gls{sub6} channel estimates are acquired from sounding reference symbols and used to predict the \gls{mmWave} \gls{DL} beam index.}
    \label{fig:what}
\end{figure}

\textbf{In this paper, we investigate} the problem of selecting the best
\gls{mmWave} \gls{DL} beam in an analog beamforming scenario, as shown
in~\cref{fig:what}. Previous work often predict the best \gls{DL} beam based on
\gls{mmWave} channel estimates, which can be difficult or costly to obtain. We
leverage the correlation between different frequency bands to learn a mapping
between the more easily acquired \gls{sub6} \gls{UL} channels and the
\gls{mmWave} \gls{DL} channels, as in~\cite{alrabeiah2019deep}. Using this
mapping, we select the best \gls{DL} beam from a pre-defined codebook. We
formulate this problem as a multi-class classification problem, where the input
is a channel estimate acquired from sounding reference symbols on the \gls{sub6}
\gls{UL} channel, and the output is a beam index used to select a column from
the codebook.
%\glsunset{IID}

\textbf{\gls{FL}}~\cite{DBLP:conf/aistats/McMahanMRHA17,DBLP:conf/mlsys/BonawitzEGHIIKK19}
has the potential to leverage decentralized datasets, such as those found in
next generation mobile networks, while enhancing privacy by using client compute
and storage resources. However, classical \gls{FL} approaches such as~\gls{FedAvg} have limitations
when data is heterogeneous and non-\gls{IID} due to the differences between
clients and between groups of clients.

\textbf{Personalized \gls{FL}} has emerged as a promising approach to address
the challenges of privacy and data heterogeneity in the context of mobile
networks, enabling mobile devices to collaboratively learn personalized models
without compromising user privacy or transmitting raw data to a central server.

Prior work~\cite{DBLP:journals/corr/abs-1910-02900,DBLP:conf/wcnc/ChafaaNBD21,
cheng2023} solved the problem with~\gls{IID} data. In this paper we use the
DeepMIMO~\cite{DBLP:journals/corr/abs-1902-06435,Remcom} dataset generation
framework to generate a more realistic dataset where the data of each client
shows non-\gls{IID} characteristics.

\textbf{Our hypothesis is} that a single \gls{FL} model cannot capture the
non-\gls{IID} characteristics present in this dataset. Using
\gls{IFCA}~\cite{DBLP:conf/nips/GhoshCYR20}, we show here that we can train a set of
global cluster models, and personalize this set to
achieve higher performance. Furthermore, using a
\gls{MoE}~\cite{DBLP:journals/corr/abs-2206-07832} we can in addition to the
global cluster models also utilize a purely local model as an expert.

In summary, our main contributions are as follows.%
\begin{enumerate}
    \def\labelenumi{\arabic{enumi}.}
    \tightlist
    \item We adapt the problem of selecting the best \gls{mmWave} analog
    beamforming \gls{DL} beam to account for non-\gls{IID} characteristics,
    specifically class imbalance, concept shift (same features, different label)
    and unbalanced data;
    \item By combining a clustering technique with a
    \gls{MoE}~\cite{DBLP:journals/corr/abs-2206-07832}
    we can adapt to non-\gls{IID} characteristics in the data to achieve higher
    accuracy than the state-of-the-art;
    \item We devise \texttt{FedLion}, a new \gls{FL} algorithm based on
    \gls{Lion}~\cite{chen2023symbolic} as the optimization algorithm on the
    parameter server which achieves higher sample efficiency and accuracy
    than~\glsxtrshort{FedAvg}.
\end{enumerate}

\section{Background}

\subsection{Analog beamforming and beam management}
With the use of \gls{mmWave} frequency bands in \glsxtrshort{5G} \gls{NR},
several \si{\giga\hertz} of bandwidth becomes available. These large amounts of
available spectrum means that very high peak rates and throughput can be
supported. However, the effective antenna area of an antenna becomes smaller
with increasing frequency for a fixed antenna gain. Hence, to maintain a given
coverage the effective antenna gain has to be increased. This can be achieved by
increasing the number of antenna elements and apply beamforming to the signal.
\begin{figure}
    \centering
    \adjustbox{width=.8\linewidth}{%
        \input{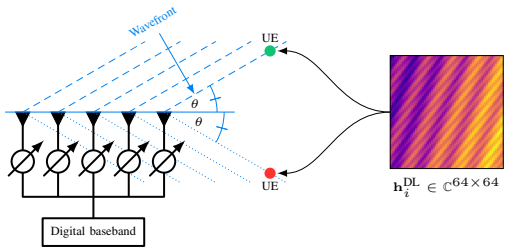}
    }
    \caption{\textbf{Antenna model.} Illustration of a \gls{ULA} antenna and a
    sample of the DL channel~\gls{hdli} between the antenna for two \glspl{UE}
    at the same distance from the antenna in \textcolor{EricssonGreen}{front of}
    or \textcolor{EricssonRed}{behind} the antenna.}
    \label{fig:ula}
\end{figure}%
Beamforming is a technique to steer the transmission (or reception) of a signal
in a specific direction, see~\cref{fig:ula} for an example with a~\glsxtrfull{ULA} used in this paper. In general, this is done by
adjusting the phase and amplitude of the transmitted signal at each antenna
element in the antenna array. In \emph{analog beamforming} the phase of the
signal fed to each antenna element is adjusted to steer the signal in the wanted
direction while the amplitude of the signal is kept constant across the antenna
elements. As the number of antenna elements increases, beams tend to become
narrower which increases the beamforming gain. Therefore, many narrow beams are
needed for coverage, which also makes it more time-consuming to find a high-gain
beam on the transmitter side.

\textbf{The problem we address in this paper} is how to select the optimal
\gls{DL} beamforming vector from a predefined codebook in order to get the best
possible user experience. We use a quantized beam steering
codebook~\cite{DBLP:journals/corr/abs-1910-02900}, \({ \gls{F} = \left[
\gls{f}(1), \ldots, \gls{f}(\gls{nobeams}) \right] }\) with \({\gls{nobeams} =
64}\) beamforming vectors as columns. Each element in~\gls{F} can be written as
\begin{equation}
    \gls{F}_{\gls{m},\gls{beamformingindex}} = \frac{1}{\sqrt{\gls{M}}} e^{-\di \pi\,m\,\cos\left.\,\gls{beamangle}\left(\gls{beamformingindex}\right)\,\right.},%
    \label{eq:codebook}
\end{equation}
 where \({\gls{m} \in \left\{0,\dots, \gls{M}{-}1\right\}}\) is the precoder weight index, for \gls{M} transmit antennas and $\gls{beamangle}(\gls{beamformingindex})= \frac{\pi \gls{beamformingindex}}{\gls{nobeams}}$ is the beam angle for beamforming vector \({
    \gls{beamformingindex} \in \left\{0,\dots, \gls{nobeams}{-}1\right\}}\).

\textbf{Traditional methods.} A simple and straightforward technique to find the
optimal \gls{DL} beam in analog beamforming is to perform an exhaustive search
over the entire set of candidate beams~\cite{Ma22}. This has large time
overheads since a \glsxtrfull{UE} will have to measure on each beam and report the
strongest. Improvements include hierarchical codebooks and
interactive beam-search.

%\subsection{Channel-beam Mapping: from sub-6GHz to \gls{mmWave}}
\textbf{Deep learning based methods}. The problem of mapping channels between frequency bands was studied
in for example~\cite{alrabeiah2019deep} for two close frequencies. The authors
of~\cite{alrabeiah2019deep} also proved the existence of a
channel-to-channel-mapping which was exploited
in~\cite{DBLP:journals/corr/abs-1910-02900} to predict \gls{mmWave} beams and
blockages from \gls{sub6} \gls{UL} channels and later
in~\cite{DBLP:conf/wcnc/ChafaaNBD21}, where \gls{FL} was used to learn the
mapping and directly predict the \gls{DL} beamforming vector. However, the
dataset used
in~\cite{DBLP:conf/wcnc/ChafaaNBD21,DBLP:journals/corr/abs-1910-02900} was
assumed to be \gls{IID} --- which is not realistic. In our work, we consider a
more realistic scenario with data having class imbalance and concept shift (same
features, different label), and we assign each data sample to the strongest base
station, see~\cref{fig:map}.

\subsection{Problem formulation}
We consider a distributed and decentralized setting with base stations as
clients \({\gls{k} \in \left\{1, 2, \ldots, \gls{K}\right\}}\). Each client
\gls{k} has access to a local data partition \gls{Pk} that never leaves the
client, and where \({\gls{nk} =\vert\gls{Pk}\vert}\) is the number of local data
samples.

We model the problem as a multi-class classification problem, where we have
\({\gls{n} = \sum_{\gls{k}=1}^{\gls{K}} \gls{nk}}\) input \gls{UL} channel
estimates~\({\gls{huli} \in \mathbb{C}^{64 \times 4}}\), indexed by~\({\gls{i} \in \left\{1, 2, \ldots,
\gls{nk}\right\}}\), and output class labels~\gls{yi} are in a finite set. See~\cite{DBLP:journals/corr/abs-1902-06435} for a detailed
expression of the channel. We further divide each client partition~\gls{Pk} into
local training and test sets and investigate the performance on the
local test set in a non-\gls{IID} setting.

Given the predicted beam index \(\gls{yihat}\), we take the corresponding
\gls{DL} beamforming vector~\(\gls{f}(\gls{yihat})\) and \gls{DL} channel
\gls{hdl} (with added \gls{AWGN}) and calculate the mean \emph{channel capacity} for a subcarrier with bandwidth~\gls{subcarrierbandwidth} as%
\begin{equation}
    %\mathcal{R}_i(\gls{hdli}, \gls{fi}) = \frac{1}{L} \sum_{l=1}^L \log_2 \left( 1 +  \text{SNR} \left\vert \gls{hdli}^\intercal{}\left[l\right] \gls{fi} \right\vert^2 \right),\label{eq:cc}
    \mathcal{R}_{\gls{i}}\left(\gls{hdli}, \gls{yihat}\right) = \frac{\gls{subcarrierbandwidth}}{\gls{L}} \sum_{\gls{subcarrierindex}=1}^{\gls{L}} \log_2 \left( 1 + \left\vert \gls{hdli}\left[\gls{subcarrierindex}\right]^{\dagger} \gls{f}(\gls{yihat}) \right\vert^2 \right),%
    \label{eq:cc}
\end{equation}
where \(\gls{hdli}\left[\gls{subcarrierindex}\right]^{\dagger}\) is the
hermitian of the \gls{DL} channel for the \(i\)th sample at the
\(\gls{subcarrierindex}\)th subcarrier and~\gls{L} is the number of subcarriers.

\subsection{\glsxtrlong{FL}}
The \gls{UL} channel estimate can be used to estimate the position of a
\gls{UE}, which is a privacy concern. One way of improving privacy is to use a
collaborative \gls{ML} algorithm such as
\gls{FedAvg}~\cite{DBLP:conf/aistats/McMahanMRHA17}. In \gls{FedAvg}, a
parameter server coordinates training of a global model in a distributed,
decentralized and synchronous manner over several communication rounds until
convergence.

In communication round \gls{t}, the parameter server selects a fraction \gls{C}
out of \gls{K} clients as the set~\gls{St}. Each selected
client~\({\gls{k}\in\gls{St}}\) trains on~\gls{nk} locally available data
samples~\({(\gls{huli}, \gls{yi}), \gls{i} \in \gls{Pk}}\), for \gls{E} epochs
before an update is sent to the parameter server. The parameter server
aggregates all received updates and computes the global model
parameters~\gls{wg}. Finally, the new global model parameters are sent to
all clients.

We can now define our learning objective as
\begin{equation}
\min_{\gls{wg} \in \R^d} \gls{flf}(\gls{wg}) \triangleq \min_{\gls{wg} \in \R^d} \overbracket[.12pt][4pt]{\sum_{k=1}^{\gls{K}} \frac{\gls{nk}}{n} \underbracket[.12pt][3pt]{\frac{1}{\gls{nk}} \sum_{i \in \gls{Pk}} \underbracket[.12pt][3pt]{\gls{l}\left(\gls{huli}, \gls{yi}, \gls{wg}\right)}_{\text{sample~\gls{i} loss}}}_{\text{client \gls{k} average loss}}}^{\text{population average loss}},%
\label{eq:loss}
\end{equation}
where \({\gls{l}\left(\gls{huli}, \gls{yi}, \gls{wg}\right)}\) is the negative
log-likelihood loss function between the optimal beam index $y_i$, i.e., the beam achieving the highest rate, and the predicted beam index $\hat{y}_i$. In other words, we aim to minimize the average loss of the global model over all clients in the population.% = \arg\max_{y\in\mathcal{F}} \mathcal{R}_{\gls{i}}\left(\gls{hdli}, y\right)$ and the 

\section{Method}

\subsection{\glsxtrlong{FedLion}}
We devise \gls{FedLion}, a federated version of
\gls{Lion}~\cite{chen2023symbolic,DBLP:conf/pldi/TilletKC19}, and use this on
the parameter server to replace \gls{FedAvg}. The pseudocode of \gls{FedLion} is
presented in \cref{alg:fl.server}, where we indicate \hl{changes} needed to
adapt \hlf{\gls{FedLion}} to work in the \gls{FL} setting. We
construct~\gls{FedLion} in a similar way to \glsxtrshort{FedAdam},
\glsxtrshort{FedYogi} and
\glsxtrshort{FedAdagrad}~\cite{DBLP:conf/iclr/ReddiCZGRKKM21} which are all
federated versions of adaptive optimizers with momentum. In addition to weight
decay (controlled by~\gls{lambda}), we also decay the server-side learning
rate~\gls{eta} as training progresses. \(\gls{beta1},\gls{beta2} \in [0,1)\) are
hyperparameters.

\begin{algorithm}[tbh]
    \begin{algorithmic}[1]
    %\Procedure{server}{$\gls{C},\gls{K}$}
        \State{}Initialize $\gls{wg}$ \Comment{Initialize global model}
        \State{}Initialize \gls{momentum} with zeros.\Comment{InitŒialize momentum}
        \State$\gls{Ks} \gets \left\lceil \gls{C}\gls{K} \right\rceil$ \Comment{Number of clients to select}
        
        \For{$\gls{t} \in \{1,2, \ldots \}$}\Comment{Until convergence}
             
            \State$\gls{St} \subseteq \gls{S}, \vert\gls{St}\vert = \gls{Ks}$\Comment{Client selection}
             
             \ForAll{$\gls{k} \in \gls{St}$}\Comment{For all clients, in parallel}
                  \State$\gls{wk}, \hl{\gls{nk}} \gets \gls{k}.\flconstfnt{client}\left(\gls{wg}\right)$\Comment{Local training}
                  \State\hl{$\gls{clientupdate} \gets \gls{wk} - \gls{wg}$}\Comment{Client  update}
            \EndFor% 

            \State\hl{$\gls{n} \gets \sum_{k \in \gls{St}} \gls{nk}$}\Comment{Total number of samples}
            \State\hl{$\gls{update} \gets \frac{1}{\gls{n}} \sum_{k \in \gls{St}} 
            \gls{nk} \gls{clientupdate}$}\Comment{Update}
            
            \State$\gls{momentum} \gets \gls{beta2} \gls{momentum} + (1-\gls{beta2})\gls{update}$\Comment{Momentum}
            \State\hlf{$\gls{wg} \gets \gls{wg} - \gls{eta}\left(\sgn{\left(\gls{beta1} \gls{momentum}  + (1-\gls{beta1})\gls{update}\right)} + \lambda \gls{wg}\right)$}
        \EndFor%
    \end{algorithmic}
    \caption{\gls{FedLion} --- server}\label{alg:fl.server}
\end{algorithm}
% \vskip-\dblfloatsep
\subsection{Personalized \glsxtrlong{FL}}
Personalized \gls{FL} offers a balance between knowledge shared among clients
and models personalized to each client. Using more than one global model
allows clusters of clients that are more similar within the cluster than to
other clients to share knowledge that is more useful to them, and to use this
as a better starting point for personalization.
\gls{IFCA}~\cite{DBLP:conf/nips/GhoshCYR20} is a clustering technique where,
after the training phase, the cluster model with the lowest loss on the
validation set is used for all future inferences. In~\gls{IFCA}, each client has
access to the full set of cluster models, and the hypothesis
of~\cite{DBLP:journals/corr/abs-2206-07832, cheng2023} is that if a client can
make use of \emph{all} of these models we can increase performance. However, in
our beam prediction dataset, we have \emph{concept shift} (same features,
different label), a non-\glsxtrshort{IID} characteristic that makes the
\gls{MoE} select a single global cluster model expert. This implies that the
conditional distribution \(\gls{P}\left(\gls{y} \middle\vert \gls{hul}\right)\)
varies between clients, but \(\gls{P}\left(\gls{hul}\right)\) is
shared~\cite{DBLP:journals/corr/abs-1912-04977}. Therefore, our \gls{MoE} use a
local model and the best global cluster model as experts, see~\cref{fig:overview}. Due to the presence of
heterogeneous environments between the clients, we also expect a \emph{concept
drift} (same label, different features) non-\glsxtrshort{IID} characteristic,
i.e. the conditional distribution \(\gls{P}\left(\gls{hul} \middle\vert
\gls{y}\right)\) varies between clients at least within the clusters, but
\(\gls{P}\left(\gls{y}\right)\) is shared.

\subsection{Models}
We take the complex \gls{UL} channel estimate~\gls{hul} as input to a \gls{CNN}
model with architecture as seen in~\cref{fig:cnn}. Since the input~\gls{hul} is
complex, we divide it into two feature channels (real and imaginary). More
precisely, we use the complex \gls{DL} channel estimate~\gls{hdl} and the
predicted beam index \(\gls{yihat}\) to calculate the channel capacity
in~\eqref{eq:cc}.

\section{Experiments}

\subsection{Dataset generation}
\begin{figure*}
    \centering
    \includegraphics[trim={2cm 0.45cm 2cm 0.2cm},clip, width=.9\textwidth]{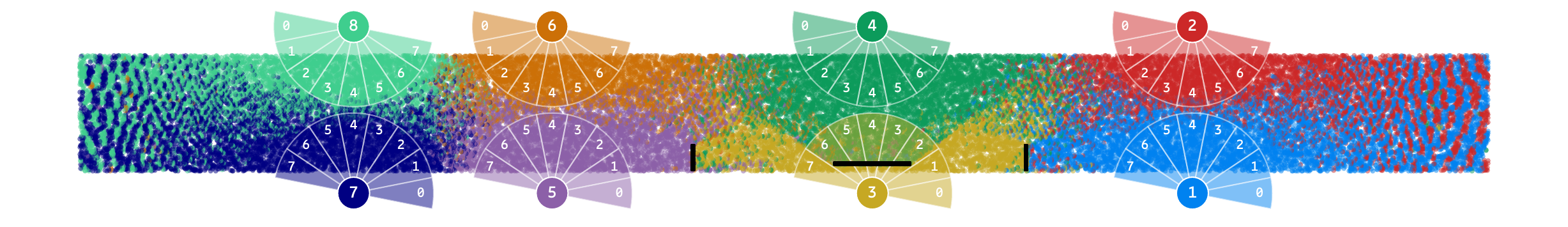}
    \caption{\textbf{Dataset.} We use DeepMimo~\cite{DBLP:journals/corr/abs-1902-06435,Remcom} to generate the \gls{mmWave} \gls{DL} channels and \gls{sub6} \gls{UL} channels for each of the 8~base stations. 
    Each position is associated with the base station that has the strongest received signal in the \gls{DL}. \emph{Note the order of the beam indicies,} here illustrated with \num{8} beams.}
    \label{fig:map}
    % \ifdefined\isaccepted%
    % \vskip-\dblfloatsep%
    % \fi
\end{figure*}

We use the publically available dataset generation framework
DeepMIMO~\cite{DBLP:journals/corr/abs-1902-06435}, which uses the Wireless
InSite ray-tracing simulator~\cite{Remcom} to generate the channels between the
\glspl{UE} and each base station. Specifically, we use the Outdoor~\num{1}
Blockage scenario where we utilize base stations on either side of the main
street as clients in the \gls{FL} terminology. For each \gls{UE} position in the
map, see~\cref{fig:map}, we generate an \gls{UL} channel on a \gls{sub6}
frequency band and a \gls{DL} channel on a \gls{mmWave} frequency band. Each
\gls{UE} position is in turn associated with the base station (client) for which
the received \gls{DL} signal is the strongest --- this results in an unbalanced
dataset. For each client, we set aside \SI{20}{\percent} of the client data as a
test set for evaluation. For evaluation, we sample \num{5000} data samples with
replacement from the test set in each run of the experiment.

In a decentralized setting it is common to have non-\gls{IID} data that can be
of non-identical client
distributions~\cite{DBLP:conf/icml/HsiehPMG20,DBLP:journals/corr/abs-1912-04977,DBLP:journals/corr/abs-2206-07832}.
In our case we generate data which can be characterized as having \emph{quantity
skew} (unbalancedness) and \emph{concept shift} (same features, different
label)~\cite{DBLP:journals/corr/abs-1912-04977}.

\begin{figure*}
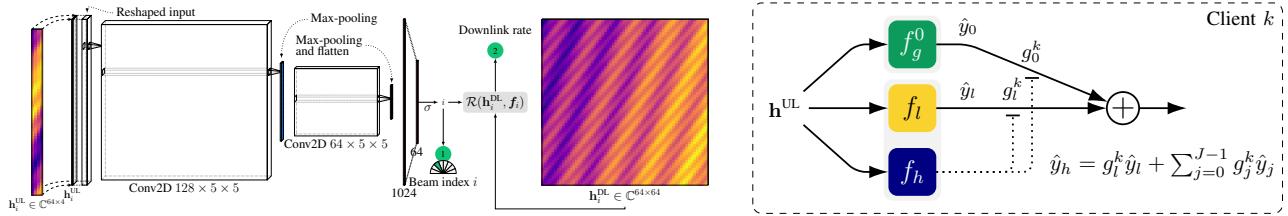

    \centering%
    \subfloat[\textbf{Neural Network architecture.} The complex \gls{UL} channel estimate~\gls{hul} is divided into two channels (real and imaginary) used as input to a \gls{CNN} model. The \gls{DL} channel estimate~\gls{hdl} is used to calculate the channel capacity.\label{fig:cnn}]{\adjustbox{height=80pt, valign=b}{\input{cnnmodel2.fig}}}%
    \hspace{.3cm}%
    \subfloat[\textbf{Adaptive expert models.} \vphantom{\gls{hul}}Our approach adjusts to non-\glsxtrshort{IID} data distributions by adaptively training a \glsxtrfull{MoE} for clients that share similar data distributions.\label{fig:overview}]{\adjustbox{height=80pt, valign=b}{\input{overview.fig}}}%
    \caption{Combining \gls{CNN}-models and a \gls{MoE} is key to good performance when data shows non-IID characteristics.}%
    % \ifdefined\isaccepted%
    % \vskip-\dblfloatsep%
    % \fi
    \end{figure*}

\subsection{Hyperparameters}
Hyperparameters are tuned using~Ray Tune\cite{DBLP:journals/corr/abs-1807-05118} in four
stages and used for all clients. For each model we tune the learning
rate~\gls{eta}, the number of hidden units in the fully connected layer,
dropout, weight decay and learning rate decay rate. Additionally, for the
\gls{IFCA} case we tune \gls{J}, \(\lambda\)~\cite{chen2023symbolic}, learning
rate, \gls{eps}-greediness~\cite{DBLP:journals/corr/abs-2206-07832} and server
learning rate decay.

First, we tune the hyperparameters for a local model and for the case with two
global cluster models \(\gls{J}=2\). Thereafter, we tune the hyperparameters for
the gating model using the best hyperparameters found in the earlier steps. Note
that we use the \emph{best} global cluster model and a local model as experts in
the \gls{MoE}. See~\cref{fig:overview}
and~\cite{DBLP:journals/corr/abs-2206-07832} for details.

Hyperparameters depend on the parameters of the data generation, but we tune the
hyperparameters for the fixed case. The tuned hyperparameters are then used for
all experiments. %See~\cref{tbl:params} for tuned hyperparameters.

\subsection{Results}
\begin{figure*}[!htb]
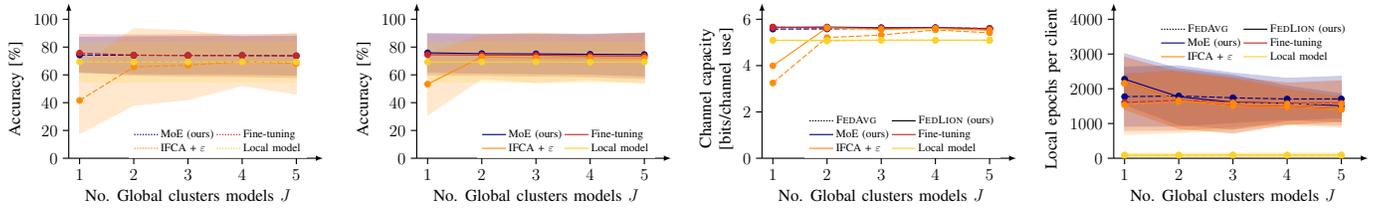

    \centering
    \subfloat[Accuracy versus number of clusters for \gls{FedAvg}.\label{fig:accvsclustersfedavg}]{\adjustbox{height=2.8cm}{\input{acc_vs_clusters_fedavg.fig}}}
    \hfill%
    \subfloat[Accuracy versus number of clusters for \gls{FedLion}.\label{fig:accvsclusterslion}]{\begin{adjustbox}{height=2.8cm}\input{acc_vs_clusters_lion.fig}\end{adjustbox}}
    \hfill%
    \subfloat[Channel capacity versus number of clusters for \gls{FedAvg} and \gls{FedLion}.\label{fig:ccvsclusters}]{\begin{adjustbox}{height=2.8cm}\input{cc_vs_clusters.fig}\end{adjustbox}}
    \hfill%
    \subfloat[Complexity in terms of local epochs versus number of clusters.\label{fig:workvsclusters}]{\begin{adjustbox}{height=2.8cm}\input{work_vs_clusters.fig}\end{adjustbox}}
\caption{\textbf{The optimal number of global cluster models.} Performance of \gls{IFCA} + $\varepsilon$, a model fine-tuned from the best global cluster model and our \gls{MoE} when varying the number of clusters \gls{J} for \gls{FedAvg} and \gls{FedLion}. The fraction of training data samples per client was \SI{10}{\percent}.}
\label{fig:clusters}%
% \ifdefined\isaccepted%
% \vskip-\dblfloatsep%
% \fi
\end{figure*}

\textbf{The optimal number of global cluster models.}
In~\cref{fig:accvsclustersfedavg,fig:accvsclusterslion,tbl:accvsclusters} we
explore the claim from~\cite{DBLP:conf/nips/GhoshCYR20} that \gls{IFCA} is
robust against setting the number of \(\gls{J}\) cluster models to be larger
than the anticipated clusters in the dataset. We set the fraction of training
data samples per client to~\SI{10}{\percent}. By utilizing the
\(\varepsilon\)-greedy cluster assignment
from~\cite{DBLP:journals/corr/abs-2206-07832} we avoid the mode-collapse problem
of vanilla \gls{IFCA}. We show that \gls{FedAvg} fails to outperform a local
model, while \gls{FedLion} does,
see~\cref{fig:accvsclustersfedavg,fig:accvsclusterslion}. 

We note that for \(\gls{J} > 2\) we get approximately the same accuracy
and channel capacity, see~\cref{fig:ccvsclusters}, while the cost of storage and
communication costs increases linearly. In~\cref{fig:workvsclusters} we see that
the convergence is faster when clients are mapped to more suitable global
cluster models. We set \(\gls{J}=2\) for the remainder of the experiments. Note
that the number of clients \(\gls{K} = 8\) is small and the fraction of
participating clients \(\gls{C}\) is \num{1}.

\begin{table*}[tbh]
    \centering
    \caption{\textbf{The optimal number of global cluster models.} Accuracy versus number of clusters for \gls{FedAvg} and \gls{FedLion}.}\label{tbl:accvsclusters}
    \begin{small}
        \begin{tabular}{c@{\hskip 4pt}lcS[table-format = 2.2]S[table-format = 2.2]S[table-format = 2.2]S[table-format = 2.2]S[table-format = 2.2]S[table-format = 2.2]S[table-format = 2.2]S[table-format = 2.2]}
\toprule
&  &         & \multicolumn{2}{c}{\textbf{MoE}} & \multicolumn{2}{c}{\textbf{Local}} & \multicolumn{2}{c}{\textbf{Fine-tuned}} & \multicolumn{2}{c}{\textbf{IFCA} + $\varepsilon$} \\
\cmidrule(lr){4-5}\cmidrule(lr){6-7}\cmidrule(lr){8-9}\cmidrule(lr){10-11}
\textbf{Clusters}&\textbf{Method}& \# trials &                                   $\mu$ &                                 $\sigma$ &          $\mu$ &  $\sigma$ &               $\mu$ &  $\sigma$ &                                     $\mu$ &                                  $\sigma$ \\
\midrule
\multirow{2}{*}{1} & \gls{FedAvg} &        29 &  \color{EricssonBlue1}74.12594834689436 &  \color{EricssonBlue1}6.5565425096472065 &      69.296724 &  7.296248 &           75.489483 &  6.819490 &   \color{EricssonGreen1}41.07577584118679 &  \color{EricssonGreen1}12.367579663218109 \\
  & \gls{FedLion} (ours) &        43 &     \color{EricssonBlue1}\textbf{75.87} &  \color{EricssonBlue1}7.1184884048215995 &      69.283488 &  7.271296 &           74.519244 &  7.401863 &  \color{EricssonGreen1}53.332674531049506 &  \color{EricssonGreen1}11.488516190827534 \\
\hdashline%
\multirow{2}{*}{2} & \gls{FedAvg} &        29 &  \color{EricssonBlue1}73.83249986582788 &   \color{EricssonBlue1}6.773198484500389 &      69.113276 &  7.283791 &           73.978793 &  7.343555 &   \color{EricssonGreen1}64.84965526646582 &  \color{EricssonGreen1}15.187930924489649 \\
  & \gls{FedLion} (ours) &        42 &     \color{EricssonBlue1}\textbf{75.28} &   \color{EricssonBlue1}7.217702923983942 &      69.251548 &  7.309131 &           74.252441 &  7.496273 &   \color{EricssonGreen1}72.88952384676251 &  \color{EricssonGreen1}7.9411447650420515 \\
\hdashline%
\multirow{2}{*}{3} & \gls{FedAvg} &        27 &  \color{EricssonBlue1}73.90944456171107 &   \color{EricssonBlue1}6.894796528252811 &      69.334444 &  7.332821 &           74.030834 &  7.733238 &    \color{EricssonGreen1}65.0429629749722 &  \color{EricssonGreen1}13.881202506470316 \\
  & \gls{FedLion} (ours) &        42 &     \color{EricssonBlue1}\textbf{75.12} &   \color{EricssonBlue1}7.239380684988816 &      69.389048 &  7.335685 &           74.165476 &  7.603039 &   \color{EricssonGreen1}71.96160717805226 &   \color{EricssonGreen1}9.398166220482052 \\
\hdashline%
\multirow{2}{*}{4} & \gls{FedAvg} &        28 &  \color{EricssonBlue1}74.03133938993726 &   \color{EricssonBlue1}6.994166168366694 &      69.337322 &  7.396636 &           73.880357 &  7.739608 &   \color{EricssonGreen1}69.54437489168984 &   \color{EricssonGreen1}8.721945656049426 \\
  & \gls{FedLion} (ours) &        41 &     \color{EricssonBlue1}\textbf{74.76} &   \color{EricssonBlue1}7.402684877572357 &      69.212622 &  7.302077 &           73.726037 &  7.722595 &   \color{EricssonGreen1}72.47829260477205 &   \color{EricssonGreen1}8.173627682765018 \\
\bottomrule
\end{tabular}

    \end{small}
    % \ifdefined\isaccepted%
    % \vskip-\dblfloatsep%
    % \fi
\end{table*}

\begin{figure*}[!htb]
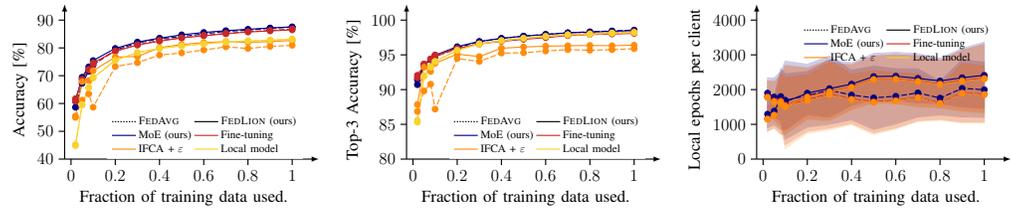

    \centering
    \subfloat[Accuracy versus fraction of training samples.\label{fig:accvstrainfrac}]{\begin{adjustbox}{height=2.8cm}\input{acc_vs_train_frac.fig}\end{adjustbox}}
    % \hfill%
    % \subfloat[Accuracy versus number of clusters for fine-tuned and \gls{MoE}.\label{fig:trainingcurves1}]{ \includegraphics[width=0.24\linewidth]{figures/acc_vs_clusters_zoom.pdf}}
    \hspace{.25cm}%
    \subfloat[Top-3 accuracy versus fraction of training samples.\label{fig:accvstrainfractop3}]{\begin{adjustbox}{height=2.8cm}\input{top3_acc_vs_train_frac.fig}\end{adjustbox}}
    \hspace{.25cm}%
    \subfloat[Complexity as local epochs versus fraction of training samples.\label{fig:workvstrainfrac}]{\begin{adjustbox}{height=2.8cm}\input{work_vs_train_frac.fig}\end{adjustbox}}
\caption{\textbf{The effect of increasing number of training data samples.} Performance when varying the fraction of training samples for \gls{FedAvg} and \gls{FedLion}.}%
\label{fig:samples}%
% \vskip-\dblfloatsep
\end{figure*}

\textbf{The effect of increasing number of training data samples.}
In~\cref{fig:accvstrainfrac} we explore the effect of increasing the number of data samples on accuracy. We can easily see that our \gls{FedLion} approach is more sample efficient than the vanilla \gls{FedAvg}, i.e.\ the algorithm attains the same accuracy using fewer samples. For the same number of samples \gls{FedLion}, achieves a higher accuracy than \gls{FedAvg}. This result carries over to the top-3 accuracy seen in~\cref{fig:accvstrainfractop3} and the effect is most prominent when we have few training data samples. Finally, we note that while \gls{FedLion} converges faster for a low fraction of training data used, \gls{FedAvg} converges faster when more than \SI{10}{\percent} of the training data is used per client, as shown in~\cref{fig:workvstrainfrac}.

\begin{figure*}[!htb]
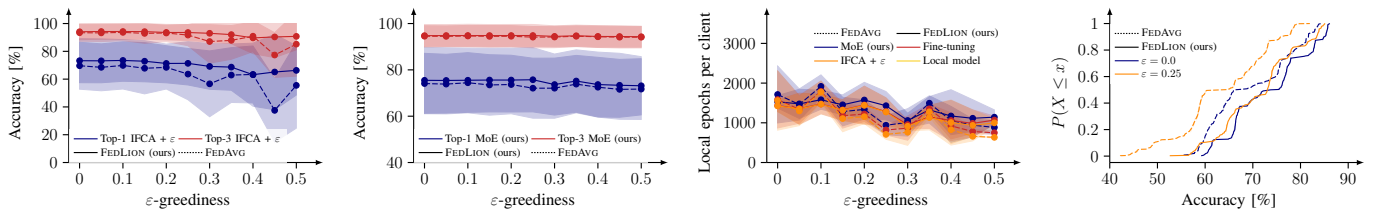

    \centering
    \subfloat[Accuracy versus \(\varepsilon\) for \gls{IFCA}.\label{fig:accvsepsifca}]{\begin{adjustbox}{height=2.8cm}\input{acc_vs_eps_ifca.fig}\end{adjustbox}}
    % \hfill%
    % \subfloat[Accuracy versus number of clusters for fine-tuned and \gls{MoE}.\label{fig:trainingcurves1}]{ \includegraphics[width=0.24\linewidth]{figures/acc_vs_clusters_zoom.pdf}}
    \hfill%
    \subfloat[Accuracy versus \(\varepsilon\) for \gls{MoE}.\label{fig:accvsepsmoe}]{\begin{adjustbox}{height=2.8cm}\input{acc_vs_eps_moe.fig}\end{adjustbox}}
    \hfill%
    \subfloat[Complexity as local epochs vs. \(\varepsilon\).\label{fig:workvseps}]{\begin{adjustbox}{height=2.8cm}\input{work_vs_eps.fig}\end{adjustbox}}
    \hfill%
    \subfloat[Accuracy \glsfmtshort{ECDF}.\label{fig:ecdfaccvseps}]{\begin{adjustbox}{height=2.8cm}\input{ecdf_acc_vs_eps.fig}\end{adjustbox}}
\caption{\textbf{Varying \(\varepsilon\)-greediness.} Performance when varying the \(\varepsilon\)-greediness in the \gls{IFCA} cluster assignment for \gls{FedAvg} and \gls{FedLion}.}%
\label{fig:eps}%
% \ifdefined\isaccepted%
% \vskip-\dblfloatsep%
% \fi
\end{figure*}

\textbf{Varying \(\varepsilon\)-greediness.}
Our~\gls{FedLion} outlined in~\cref{alg:fl.server} method is more robust w.r.t. increasing \(\varepsilon\), shown in~\cref{fig:accvsepsifca,fig:ecdfaccvseps}. The fraction of training data samples was \SI{10}{\percent}. We see in~\cref{fig:workvseps} that the convergence rate improves as \(\varepsilon\) increases. We hypothesise that this stems from the fact that the algorithm is robust against noisy updates due to its momentum term, and that this helps prevent overfitting. However, the effect on accuracy is not seen in the case of \gls{MoE}, see~\cref{fig:accvsepsmoe}. \gls{FedLion} outperforms \gls{FedLion} for all values of \(\varepsilon\).

\textbf{Impact of \glsentryfull{SNR}.}
To investigate the impact of \gls{SNR}, we set the \gls{UL} \gls{SNR} for all data samples to be the same in one run of the experiment. The fraction of training data samples was \SI{10}{\percent}. In~\cref{fig:accvssnrifca} we show the result on accuracy for \gls{IFCA} with \(\varepsilon\)-greedy cluster assignment. Our \gls{FedLion} server-side optimization method achieves slightly higher accuracy that \gls{FedAvg} for \gls{IFCA}, but the top-3 accuracy is indistinguishable between the two. For \gls{MoE} the performance is similar for \gls{FedLion} and \gls{FedAvg}.

\begin{figure}
    \begin{minipage}[t]{.49\linewidth}%
        \centering%
        \begin{adjustbox}{height=2.8 cm}\input{acc_vs_ulsnr_fl.fig}\end{adjustbox}% 
        \vspace*{-2mm}%
        \caption{\textbf{Impact of \glsentryfull{SNR}} for \gls{IFCA} with 
        \(\varepsilon\)-greedy cluster assignment.}%
        \label{fig:accvssnrifca}%
    \end{minipage}%
    \hfill%
    \begin{minipage}[t]{.49\linewidth}%
        \centering%
        \begin{adjustbox}{height=2.8 cm}\input{training_curve_0.fig}\end{adjustbox}%
        \vspace*{-2mm}%
        \caption{\textbf{Training curves} for a \gls{IFCA} global cluster
        model with \(\varepsilon\)-greedy cluster assignment. }%
        \label{fig:trainingcurves}%
    \end{minipage}%
    %\vskip-\dblfloatsep
\end{figure}

\textbf{Training curves.}
Finally, we illustrate training curves for the first of the two global cluster models in~\cref{fig:trainingcurves}. We see that initially, \gls{FedAvg} improves faster than \gls{FedLion}. However, \gls{FedLion} catches up and achieves higher performance.

\section{Related work}

\textbf{\gls{ML}-based beam alignment.} The \gls{mmWave} beam alignment problem
has received great attention in recent years~\cite{DBLP:journals/wcl/TianZXY23,
DBLP:journals/icl/TianZXY22, Echigo21}. Most data-driven approaches in this area
focused on supervised learning procedures to predict the best beam based on some
form of feature side information~\cite{DBLP:journals/corr/abs-1910-02900}. For
example, Echigo \etal.~\cite{Echigo21} proposed a deep learning approach to
predict an optimal narrow beam based on wide beams measurements, reducing the
beam alignment overhead. Alrabeiah and
Alkhateeb~\cite{DBLP:journals/corr/abs-1910-02900} used an neural network to
predict the best beam in \gls{mmWave} networks based on sub-\SI{6}{\giga\hertz}
feature information. Our problem formulation was inspired by this work, but we
devise a personalized \gls{FL} approach to deal with the inherent non-\gls{IID}
nature of the \gls{mmWave} channel. Tian
\etal.~\cite{DBLP:journals/wcl/TianZXY23} studied a beam-alignment problem in
\gls{mmWave} vehicular network. They devise a personalized deep learning
approach which envision a pre-training phase on the complete dataset and a local
fine-tuning phase on data from the specific base station. As opposed to our
\gls{FL} method, classical deep learning approaches require the training to be
centralized; as a consequence they are harder to scale and raise privacy
concerns.

\textbf{\gls{FL}-based beam alignment.} A few works investigated the beam
alignment problem in \gls{FL}~\cite{Elbir20, Chafaa21}. Elbir
\etal.~\cite{Elbir20} introduce an \gls{FL} framework for hybrid beamforming
using \glspl{CNN} which predicts the best analog beamformers, based on channel
data. Similarly, Chafaa \etal.~\cite{Chafaa21} employ an \gls{FL} method to
predict the optimal beamforming vector from a discrete beamforming codebook,
based on \gls{sub6} channel data. Although these methods have the advantage of
providing decentralized training and execution, and hence reducing communication
overhead, they lack personalization and may lead to suboptimal beamforming
configurations, especially in highly heterogeneous scenarios. To the best of our
knowledge, this paper is the first to propose a personalized \gls{FL} approach.
\section{Discussion and limitations}
There are many future directions of research stemming from this work. Firstly,
and most importantly, these methods must be tested on a real-world dataset with
more clients, where the environment is more complex and with a set of realistic
beams. A real-world dataset will also have a \emph{concept drift}
non-\glsxtrshort{IID} characteristic (same labels, different
features)~\cite{DBLP:journals/corr/abs-1912-04977} that we expect our \gls{MoE}
method to handle better than the concept shift characteristic. Secondly, the
data volume is an important for the deployment of these methods in real mobile
networks. Thirdly, the generalization aspect has not been investigated even
though it is a major advantage of the \gls{MoE} approach.

\section{Conclusions}

In this paper we investigated the problem of selecting the best \gls{mmWave}
downlink beam in an analog beamforming scenario, when data has non-\gls{IID} characteristics, specifically class imbalance, concept shift and unbalanced data --- a more difficult problem than the problem solved in prior work.

We leveraged a personalized \gls{FL} technique~\cite{DBLP:journals/corr/abs-2206-07832} that is able to adapt to non-\glsxtrshort{IID} characteristics and showed that it is robust to incorrectly setting the number of global cluster models.

In conclusion, we demonstrated higher sample-efficiency and higher accuracy than prior works by combining personalized \gls{FL} with a server-side optimization method \gls{FedLion}. Furthermore, we provided insights and opportunities for future research and practical use of these methods.

\section*{Acknowledgment}
\addcontentsline{toc}{section}{Acknowledgment}
\ifdefined\isaccepted We thank Y.\,Cheng for his inspirational Master's thesis work and all reviewers for their critical feedback that enhanced the quality of this paper, especially Dr.\,A.\,Alam, Dr.\,A.\,Alabbasi, Dr.\,N.\,Jaldén, Dr.\,J.\,Jeong, D.\,Kolmas, Dr.\,I.~Mitsioni, Dr.\,C.\,Svahn, G.\,Verardo, Assoc.\,Prof.\,Š.\,Girdzijauskas and Prof.\ S.\ Haridi.\else We thank Y.~Cheng for his inspirational Master's thesis work.
\fi 

This work was partially supported by the \gls{WASP} 
funded by the Knut and Alice Wallenberg Foundation.\bibliographystyle{IEEEtran}
\balance \bibliography{IEEEabrv,ref}\balance
\ifdefined\isarxiv \cleardoublepage \appendix
    \section{Appendix}\label{sec:appendix}
    \subsection{Hyperparameters}
    Hyperparameters used for the dataset generation are listed in~\cref{tbl:deepmimoparams} and for the training in~\cref{tbl:params_data,tbl:params_fl_fedavg,tbl:params_fl_fedlion,tbl:params_ft,tbl:params_local,tbl:params_moe_fedlion,tbl:params_moe_fedavg}.
  \begin{table}
      \caption{DeepMimo~\cite{DBLP:journals/corr/abs-1902-06435} dataset parameters.}
      \scriptsize
\begin{tabularx}{\linewidth}{Tcc}
    \toprule
    \multicolumn{1}{X}{\textbf{Parameter}}  &  \textbf{DL} & \textbf{UL} \\
    \midrule
scenario &                    \texttt{O1\_28B} &                   \texttt{O1\_3p5B} \\
num\_paths &                         5 &                        15 \\
active\_BS &  $[1, 2, \ldots, 8]$ &  $[1, 2, \ldots, 8]$ \\
user\_row\_first &                         1 &                         1 \\
user\_row\_last &                      2200 &                      2200 \\
row\_subsampling &                         1 &                         1 \\
user\_subsampling &                         1 &                         1 \\
enable\_BS2BS &                     \texttt{False} &                     \texttt{False} \\
OFDM\_channels &                         1 &                         1 \\
BS2BS\_isnumpy &                      \texttt{True} &                      \texttt{True} \\
dynamic\_settings.first\_scene &                         1 &                         1 \\
dynamic\_settings.last\_scene &                         1 &                         1 \\
bs\_antenna.shape &                [1, 64, 1] &                 [1, 4, 1] \\
bs\_antenna.spacing &                       0.5 &                       0.5 \\
bs\_antenna.radiation\_pattern &                 \texttt{isotropic} &                 \texttt{isotropic} \\
bs\_antenna.rotation &                      \texttt{None} &                      \texttt{None} \\
ue\_antenna.shape &                 [1, 1, 1] &                 [1, 1, 1] \\
ue\_antenna.spacing &                       0.5 &                       0.5 \\
ue\_antenna.radiation\_pattern &                 \texttt{isotropic} &                 \texttt{isotropic} \\
OFDM.subcarriers &                       512 &                        32 \\
OFDM.subcarriers\_limit &                        64 &                        64 \\
OFDM.subcarriers\_sampling &                         1 &                         1 \\
OFDM.bandwidth &                       \SI{0.5}{\giga\hertz} &                      \SI{0.02}{\giga\hertz} \\
OFDM.RX\_filter &                         0 &                         0 \\
scenario\_params.carrier\_freq &                   \SI{28}{\giga\hertz} &                   \SI{3.5}{\giga\hertz} \\
scenario\_params.tx\_power &                         0 &                         0 \\
scenario\_params.num\_BS &                        12 &                        12 \\
scenario\_params.user\_grids &          [[1, 2751, 181]] &          [[1, 2751, 181]] \\
    \bottomrule
    \end{tabularx}
      \label{tbl:deepmimoparams}
  \end{table}

  \begin{table}
      \caption{Training hyperparameters related to the dataset.}
      \scriptsize
      \begin{tabularx}{\linewidth}{TS[table-format=2.5e9,round-pad=false,round-mode=none]}
    \toprule
    \multicolumn{1}{X}{\textbf{Parameter}} &        \textbf{Value} \\
    \midrule
                          data.n\_data &           -1 \\
                     data.n\_data\_test &         5000 \\
                     data.num\_classes &           64 \\
                        data.channels &            2 \\
                      data.train\_frac &          0.1 \\
                data.eval\_num\_clients &            8 \\
                     data.num\_clients &            8 \\
                       data.add\_noise &      \texttt{physics} \\
                             data.snr &           80 \\
                 data.ue\_tx\_power\_dBm &           23 \\
                 data.bs\_tx\_power\_dBm &           34 \\
                 data.noise\_figure\_dB &            5 \\
                 data.interference\_dB &            0 \\
                      data.label\_type &    \texttt{realistic} \\
    \bottomrule
    \end{tabularx}
      \label{tbl:params_data}
  \end{table}

  \begin{table}
    \caption{Training hyperparameters related \gls{FL} and \gls{FedLion}.}
    \scriptsize
    \begin{tabularx}{\linewidth}{TS[table-format=2.5e9,round-pad=false,round-mode=none]}
  \toprule
  \multicolumn{1}{X}{\textbf{Parameter}} &        \textbf{Value} \\
  \midrule
federated.lr &   0.00341691 \\
                  federated.server\_lr &  0.000272589 \\
                      federated.lmbda &  0.000463701 \\
                  federated.fldropout &     0.583516 \\
            federated.fl\_weight\_decay &  2.10218e-06 \\
       federated.server\_lr\_decay\_rate &   4.6406e-05 \\
     federated.fl\_local\_lr\_decay\_rate &  1.00885e-07 \\
                        federated.eps &    0.0351229 \\
             federated.flhiddenunits1 &         1024 \\
             federated.flhiddenunits2 &         2048 \\
                     federated.epochs &         4000 \\
                federated.fl\_patience &            5 \\
                   federated.clusters &            2 \\
                       federated.frac &            1 \\
                   federated.local\_bs &          256 \\
                   federated.local\_ep &            3 \\
                 federated.flfilters1 &          128 \\
                 federated.flfilters2 &           64 \\
                 federated.filtersize          &            5 \\
               federated.server\_optim &         \texttt{FedLion} \\
                      federated.beta1 &         0.95 \\
                      federated.beta2 &         0.98 \\
                        federated.tau &        0.001 \\
    \bottomrule
    \end{tabularx}
    \label{tbl:params_fl_fedlion}
\end{table}

\begin{table}
  \caption{Training hyperparameters related \gls{FL} and \gls{FedAvg}.}
  \scriptsize
  \begin{tabularx}{\linewidth}{TS[table-format=2.5e9,round-pad=false,round-mode=none]}
  \toprule
  \multicolumn{1}{X}{\textbf{Parameter}} &        \textbf{Value} \\
  \midrule
federated.lr &    0.0179516 \\
federated.server\_lr &   0.00132305 \\
federated.fldropout &     0.728812 \\
federated.fl\_weight\_decay &  1.34272e-07 \\
federated.server\_lr\_decay\_rate &  0.000264015 \\
federated.fl\_local\_lr\_decay\_rate &   5.8123e-07 \\
federated.eps &   0.00502087 \\
federated.lmbda &  3.11622e-05 \\
federated.flhiddenunits1 &         1024 \\
federated.flhiddenunits2 &         2048 \\
federated.epochs &         4000 \\
federated.fl\_patience &            5 \\
federated.clusters &            2 \\
federated.frac &            1 \\
federated.local\_bs &          256 \\
federated.local\_ep &            3 \\
federated.flfilters1 &          128 \\
federated.flfilters2 &           64 \\
federated.filtersize          &            5 \\
federated.server\_optim &       \texttt{FedAvg} \\
federated.beta1 &         0.95 \\
federated.beta2 &         0.98 \\
federated.tau &        0.001 \\
    \bottomrule
\end{tabularx}
  \label{tbl:params_fl_fedavg}
\end{table}

  \begin{table}
    \caption{Training hyperparameters related to fine-tuning of \gls{FL} models.}
    \scriptsize
    \begin{tabularx}{\linewidth}{TS[table-format=2.5e9,round-pad=false,round-mode=none]}
  \toprule
  \multicolumn{1}{X}{\textbf{Parameter}} &        \textbf{Value} \\
  \midrule
  finetuning.ft\_lr &   0.00011723 \\
  finetuning.ft\_weight\_decay &      0.26246 \\
 finetuning.ft\_lr\_decay\_rate &     0.025521 \\
      finetuning.ft\_patience &           10 \\
    \bottomrule
    \end{tabularx}
    \label{tbl:params_ft}
\end{table}

\begin{table}
  \caption{Training hyperparameters related to \gls{MoE} models and \gls{FedLion}.}
  \scriptsize
  \begin{tabularx}{\linewidth}{TS[table-format=2.5e9,round-pad=false,round-mode=none]}
    \toprule
    \multicolumn{1}{X}{\textbf{Parameter}} &        \textbf{Value} \\
    \midrule
                           moe.moe\_lr &  5.12969e-06 \\
                     moe.gate\_dropout &     0.413556 \\
                moe.moe\_lr\_decay\_rate &   0.00079047 \\
                     moe.gatefilters1 &           16 \\
                     moe.gatefilters2 &            0 \\
                moe.gate\_weight\_decay &  2.15771e-06 \\
                       moe.moe\_epochs &          400 \\
                 moe.gatehiddenunits1 &           16 \\
                 moe.gatehiddenunits2 &            8 \\
                   moe.gatefiltersize &            5 \\
                     moe.moe\_patience &           10 \\
\bottomrule
\end{tabularx}
  \label{tbl:params_moe_fedlion}
\end{table}

\begin{table}
  \caption{Training hyperparameters related to \gls{MoE} models and \gls{FedAvg}.}
  \scriptsize
  \begin{tabularx}{\linewidth}{TS[table-format=2.5e9,round-pad=false,round-mode=none]}
    \toprule
    \multicolumn{1}{X}{\textbf{Parameter}} &        \textbf{Value} \\
    \midrule
    moe.moe\_lr                 &  8.15905e-06 \\
    moe.gate\_dropout           &     0.554633 \\
    moe.gatehiddenunits1        &            4 \\
    moe.moe\_lr\_decay\_rate    &  0.000202676 \\
    moe.gatefilters1            &           16 \\
    moe.gatefilters2            &            4 \\
    moe.gate\_weight\_decay     &  4.23931e-06 \\
    moe.moe\_epochs             &          400 \\
    moe.gatehiddenunits2        &            8 \\
    moe.gatefiltersize          &            5 \\
    moe.moe\_patience           &           10 \\
    \bottomrule
\end{tabularx}
  \label{tbl:params_moe_fedavg}
\end{table}

\begin{table}
  \caption{Training hyperparameters related to local models.}
  \scriptsize
  \begin{tabularx}{\linewidth}{TS[table-format=2.5e9,round-pad=false,round-mode=none]}
    \toprule
    \multicolumn{1}{X}{\textbf{Parameter}} &        \textbf{Value} \\
    \midrule
                     local.loc\_epochs &          400 \\
                       local.local\_lr &   0.00109789 \\
             local.local\_weight\_decay &    0.0236955 \\
                   local.localdropout &     0.694774 \\
            local.local\_lr\_decay\_rate &    0.0214379 \\
              local.localhiddenunits1 &         1024 \\
              local.localhiddenunits2 &         2048 \\
                  local.localfilters1 &          128 \\
                  local.localfilters2 &           64 \\
                  local.filtersize          &            5 \\
                 local.local\_patience &           10 \\
    \bottomrule
\end{tabularx}
  \label{tbl:params_local}
\end{table}

    \clearpage
    \printglossary[type=acronym,style=index]\printglossary[type=symbols,style=index]\fi
\end{document}